\documentclass[twocolumn,twoside,slac_two]{revtex4}
\usepackage{graphicx}
\usepackage{fancyhdr}
\begin{document}

\title{The Galactic Center Region Imaged by VERITAS}


\author{M.~Beilicke\footnote{E-mail: beilicke@physics.wustl.edu}}
\affiliation{Department of Physics and McDonnell Center for the Space
Sciences, Washington University in St.~Louis, 1 Brookings Drive,
St.~Louis, MO 63130, USA}

\author{for the VERITAS Collaboration}
\affiliation{{\tt http://veritas.sao.arizona.edu}}

%% ############################################################
%% ############ Abstract
%% ############################################################
\begin{abstract}

The Galactic Center has long been a region of interest for high-energy
and very-high-energy observations. Many potential sources of GeV/TeV
$\gamma$-ray emission have been suggested, e.g., the accretion of matter
onto the black hole, cosmic rays from a nearby supernova remnant, or the
annihilation of dark matter particles. The Galactic Center has been
detected at MeV/GeV energies by EGRET and recently by Fermi/LAT. At
GeV/TeV energies, the Galactic Center was detected by different
ground-based Cherenkov telescopes such as CANGAROO, Whipple $10 \,
\rm{m}$, H.E.S.S., and MAGIC. We present the results from $15 \,
\rm{hrs}$ of VERITAS observations conducted at large zenith angles,
resulting in a $>$$10$ standard deviation detection and confirmation of
the high-energy spectrum observed by H.E.S.S. The combined Fermi/VERITAS
results are compared to astrophysical models.

\end{abstract}

\maketitle

\thispagestyle{fancy}


%% ############################################################
%% ############ Introduction
%% ############################################################
\section{Introduction}
\label{sec:Introduction}

%% ############################################################
\subsection{The Galactic Center region}

The center of our galaxy harbors a $4 \times 10^{6} \, M_{\odot}$ black
hole (BH) which is believed to coincide with the strong radio source
Sgr\,A*. At optical wavelengths the view towards the Galactic Center
(GC) is hidden by molecular clouds and dust. X-ray transients with $2-10
\, \rm{keV}$ energy output up to $10^{35} \, \rm{ergs}/s$ are observed
on a regular basis, as well as transients at MeV/GeV
energies\footnote{See for example the recent Astronomer's telegrams
\#2690, \#2770, \#2770, \#3123, \#3162, \#3163, \#3183}. Besides these
transients, there are other astrophysical sources located in the close
vicinity of the GC which may potentially be capable of accelerating
particles to multi-TeV energies, such as the supernova remnant
Sgr\,A~East or a plerion found in that region \cite{GC_Plerion}.

The gravitational potential of our galaxy is believed to bind a halo of
dark matter particles~-- the nature of which is still a matter of very
active research. The super-symmetric neutralinos $\chi$ are discussed as
one potential dark matter particle accumulating in this halo~-- the
density of which peaks at the GC. Neutralinos could annihilate directly
to gamma rays forming narrow lines (through $\chi \chi \rightarrow
\gamma \gamma$ or $\chi \chi \rightarrow \gamma + Z^{0}$) or annihilate
to quarks or heavy leptons, hadronizing and producing secondary $\gamma$
rays in a continuum \cite{Neutralino}. The resulting spectrum would have
a cut-off near the neutralino mass $m_{\chi}$, and a detailed spectral
shape determined by the annihilation channel. The resulting signal would
typically show up in the GeV/TeV band for natural neutralino parameters.
Assuming a certain density profile of the dark matter the expected
$\gamma$-ray flux along the line-of-sight integral can be calculated as
a function of $m_{\chi}$ and the annihilation cross section
\cite{GammasFromNWF} and can in turn be compared to measurements or
upper limits.

%% ############################################################
\subsection{The Galactic Center seen at GeV/TeV energies}

The GC region is crowded with astrophysical sources which can
potentially emit $\gamma$-rays at MeV/GeV/TeV energies. The limited
resolution of instruments in these wave bands makes definite
associations challenging. The EGRET $\gamma$-ray telescope detected a
MeV/GeV source 3EG\,J1746-2851 which is spatially coincident with the GC
\cite{Egret_GC}. Recently, the Fermi/LAT resolved more than one MeV/GeV
sources in the GC region \cite{Fermi_FirstCatalog}, where the strongest
source is spatially coincident with the GC (Fig.~\ref{fig:Skymap}).
However, uncertainties in the diffuse galactic background models and the
limited angular resolution of the Fermi/LAT make it difficult to study
the morphologies of these MeV/GeV sources.

At GeV/TeV energies a detection from the direction of the GC was first
reported in 2001/02 by the CANGAROO\,II collaboration which operated a
ground-based $\gamma$-ray telescope. A steep energy spectrum
$\rm{d}N/\rm{d}E \propto E^{-4.6}$ was reported with an integral flux at
the level of $10\%$ of the Crab Nebula flux \cite{CANGAROO_GC}. Shortly
after, evidence at the level of $3.7$ standard deviations (s.d.) was
reported from 1995-2003 observation conducted at large zenith angles
(LZA) with the Whipple $10 \, \rm{m}$ $\gamma$-ray telescope
\cite{Whipple_GC}.

%-------------
\begin{figure*}[t]
 \includegraphics[width=0.44\textwidth]{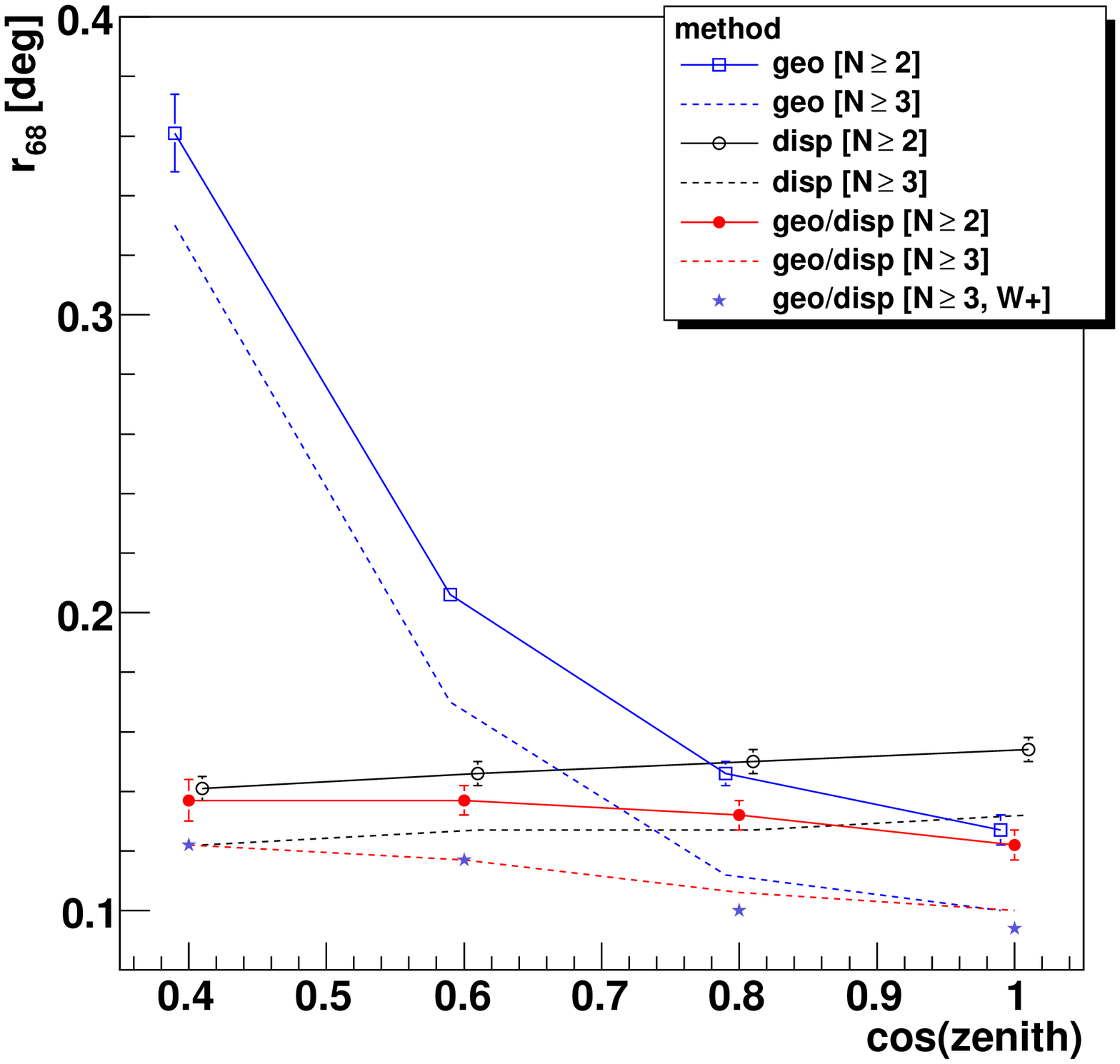}
 \hfill
 \includegraphics[width=0.55\textwidth]{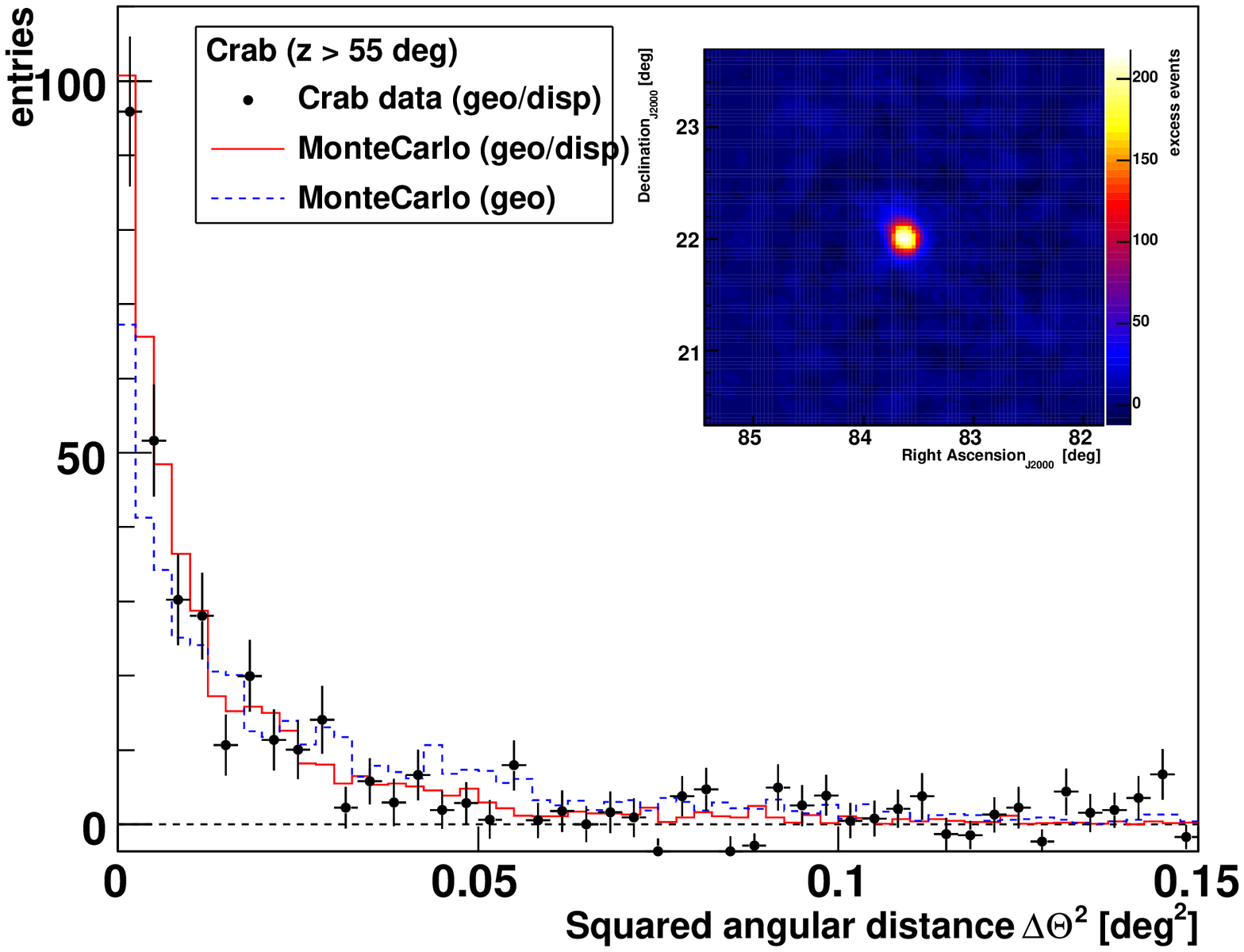}

\caption{\label{fig:LZA} {\bf Left:} VERITAS angular resolution
($r_{68}$ containment radius) as a function of $\cos(z)$. The {\it geo}
algorithm performs well for zenith angles $<40 \deg$ ($\cos(z) = 0.8$)
but gets worse below. The {\it disp} algorithm does not depend on the
stereo angle between image axis and is therefore not sensitive to the
zenith angle.  At LZA of $65 \, \deg$ it outcompetes the {\it geo}
algorithm by a factor of more than $2$. A weighted combination of both
algorithms ({\it geo}/{\it disp}, see text) gives an almost flat angular
resolution.  {\bf Right:} The data points show the angular distribution
of excess events from $3.5 \, \rm{hrs}$ of Crab Nebula observations
taken at zenith angles $z>55 \, \deg$. The showers were reconstructed
with the {\it geo}/{\it disp} method. The solid line represents the
angular distribution of MonteCarlo events ({\it geo}/{\it disp} method)
covering the same zenith angle range as the data. The dashed line shows
the distribution of MonteCarlo events which were reconstructed with the
standard {\it geo} algorithm. The inlay shows the corresponding sky map
of the Crab Nebula data.}

\end{figure*}

The GC was finally confirmed as a GeV/TeV $\gamma$-ray source in a
highly significant ($>$$60$ s.d.) detection from 2004-2006 observations
reported by the H.E.S.S. collaboration \cite{HESS_SgrA}.  The energy
spectrum was well described by a power-law $\rm{d}N/\rm{d}E \propto
E^{-2.1}$ with a cut-off at $\sim$$15 \, \rm{TeV}$. No evidence for
variability was found in the H.E.S.S. or Whipple data over a time scale
of more than 10 years. Using a high-precision pointing system of the
H.E.S.S. telescopes the position of the supernova remnant Sgr\,A~East
could be excluded as the source of the $\gamma$-ray emission. After
subtracting the point source located at the position of the GC the
H.E.S.S. collaboration was able to identify a diffuse GeV/TeV
$\gamma$-ray emission. The intensity profile of the diffuse component is
found to be aligned along the galactic plane and follows the structure
of molecular clouds \cite{HESS_SgrA_Diffuse}. The energy spectrum of the
diffuse emission (dashed contour lines in Fig.~\ref{fig:Skymap}) can be
described by a power-law $\rm{d}N/\rm{d}E \propto E^{-2.3}$ and was
explained by an interaction of local cosmic rays (CRs) with the matter
in the molecular clouds~-- indicating a harder spectrum and a higher
flux of CRs in this region as compared to the CRs observed at Earth.

The MAGIC collaboration detected the GC in 2004/05 observations
performed at LZA at the level of $7$ s.d. \cite{MAGIC_GC}, confirming
the energy spectrum measured by H.E.S.S. The differences between the
energy spectrum measured by CANGAROO compared to the spectra measured by
the other ground-based GeV/TeV instruments could perhaps be explained if
the different instruments observed different astrophysical sources.

%% ############################################################
%% ############ LZA observations
%% ############################################################
\section{Large zenith-angle observations}
\label{sec:LZA}

The standard method of shower reconstruction in arrays of ground-based
Cherenkov telescopes (such as VERITAS) is based on the intersection of
the major axis of the Hillas images recorded in the individual
telescopes \cite{Hofmann1999}. This stereoscopic method is generally
very powerful, since it makes use of the full capabilities of the
stereoscopic recording of showers. In the following this method is
referred to as {\it geo} (geometrical) method.

An alternative technique has been developed long ago for data taken with
single-telescopes (i.e. Whipple $10 \, \rm{m}$), using an estimate of
the {\it displacement} parameter which is measured between the center of
gravity (CoG) of the Hillas ellipse and the shower position in the
camera system \cite{BuckelyDisp}. For $\gamma$-ray showers the {\it
displacement} parameter has a certain characteristic value as a function
of the image parameters. The characteristic displacement can be
parameterized as a function of the length $l$, the width $w$, and the
amplitude/size $s$ of the corresponding image. Throughout this paper
this method is referred to as {\it disp} method.

In LZA observations the telescope's locations in the plane perpendicular
to the shower axis are 'shrinking' towards one dimension (due to
projection effects); this strongly reduces the average stereo angle
between the major axes of pairs of images, causing a large uncertainty
in the determination of the intersection point. This, in turn, leads to
a considerable reduction of the angular resolution in the reconstruction
of the shower direction and impact parameter. The {\it disp} method, on
the other hand, does not rely on the intersection of axes, making it
independent of the stereo angle between images. Therefore, no
substantial drop in performance is expected with increasing zenith
angle. The {\it disp} parameter was implemented into the VERITAS
analysis chain being parameterized as a function of $l$, $w$, $s$, the
zenith angle $z$, the azimuth angle $Az$, as well as the pedestal
variance of the image. For each image the {\it disp} parameter is read
from a 6-dimensional look-up table which was trained using MonteCarlo
simulations. For each image the corresponding disp parameter results in
two most likely points of the shower direction (camera coordinates):
$\rm{CoG} \pm \rm{\it disp}$ along the major axis of the parameterized
image. The two-fold ambiguity is resolved by combining the points of all
images involved in the event. The shower impact parameter is
reconstructed in a comparable way.

Figure~\ref{fig:LZA}, left shows the angular resolution of both methods
({\it geo} and {\it disp}) as a function of the cosine of the zenith
angle $z$.  While the {\it disp} method remains almost independent of
$\cos (z)$, the resolution of the standard method {\it geo} becomes
increasingly worse at LZA. A further improvement is achieved if both
methods are combined: $d = d_{\rm{geo}} \cdot (1-w') + d_{\rm{disp}}
\cdot w'$. The weight is calculated as $w' = \exp(-12.5 \cdot
(\cos(z)-0.4)^{2})$; for $\cos(z) < 0.4$ the weight is set to $w' = 1$.
The method was tested on Crab Nebula data taken at LZA
(Figure~\ref{fig:LZA}, right). The data are in excellent agreement with
the simulations and illustrate the clear improvement the {\it disp}
method provides in the case of LZA observations.

%% ############################################################
%% ############ VERITAS observations of the galactic center
%% ############################################################
\section{The Galactic Center region imaged by VERITAS}

%% ############################################################
\subsection{VERITAS observations}

%-------------
\begin{figure}[t]

  \includegraphics[width=0.49\textwidth]{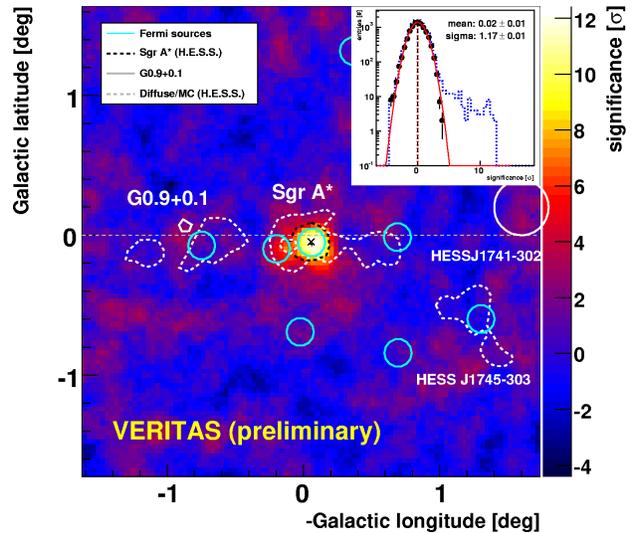}

\caption{\label{fig:Skymap} VERITAS sky map of the GC region (excess
significances, ring background, smoothed by $0.12 \deg$).  The positions
of the excess fit (cross) and the GC ('x') are shown, as well. The black
dashed contour lines indicate the GC as seen by H.E.S.S.
\cite{HESS_SgrA} and the gray solid line indicates the excess from the
supernova remnant G\,0.9+0.1 (H.E.S.S.). The gray dashed lines indicate
the H.E.S.S. diffuse emission along the galactic plane and from
HESS\,J1745-303 \cite{HESS_SgrA_Diffuse}. The position of
HESS\,J1741-302 is indicated, as well (circle). The solid circles (cyan
color) indicate the positions of the MeV/GeV sources taken from the
first Fermi catalog \cite{Fermi_FirstCatalog}. The inlay shows the
distribution of significances from the VERITAS sky map including (dashed
line) and excluding (data points) the GC region.}

\end{figure}

%-------------
\begin{figure*}[t]

\includegraphics[width=0.49\textwidth]{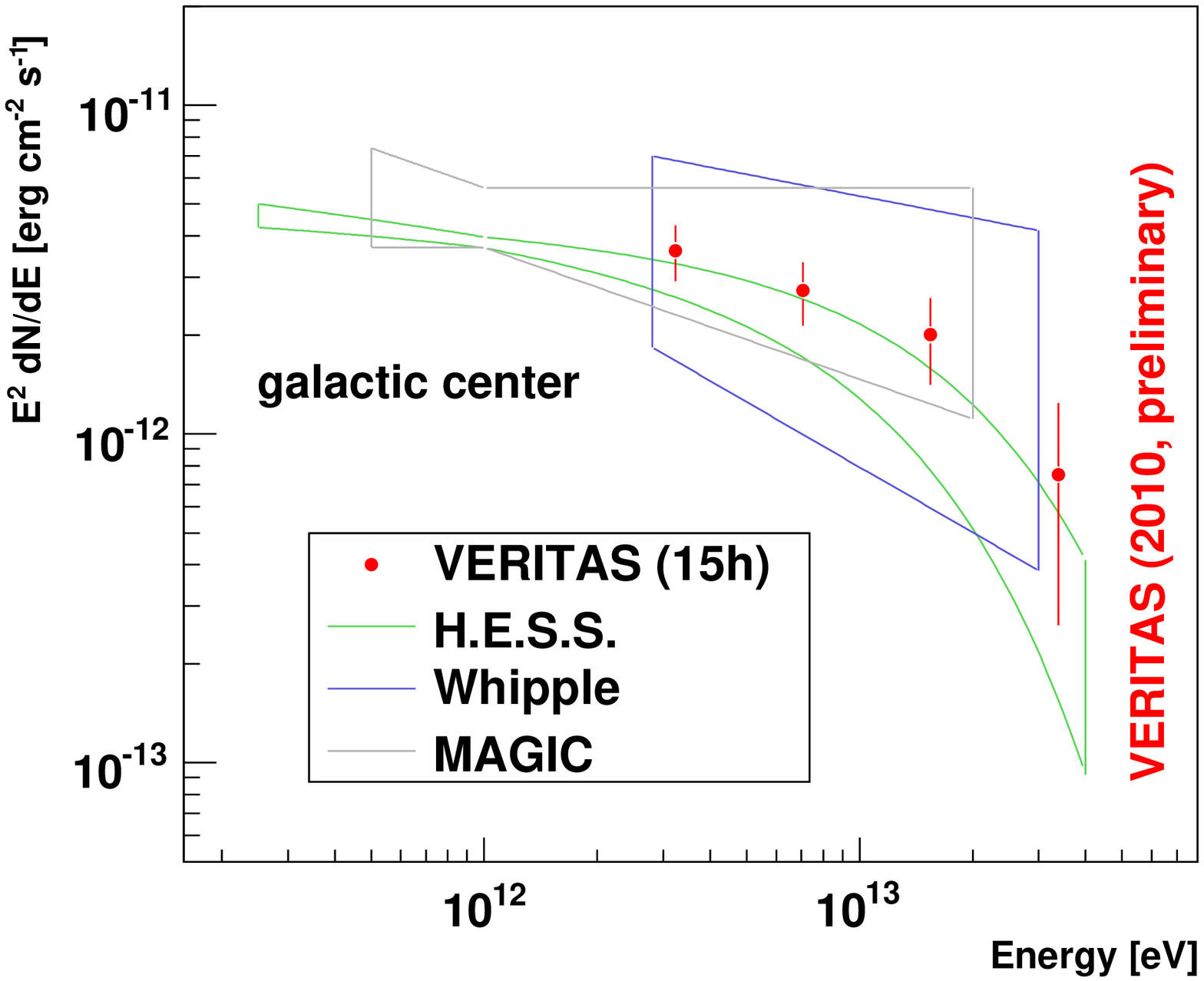}
\hfill
\includegraphics[width=0.49\textwidth]{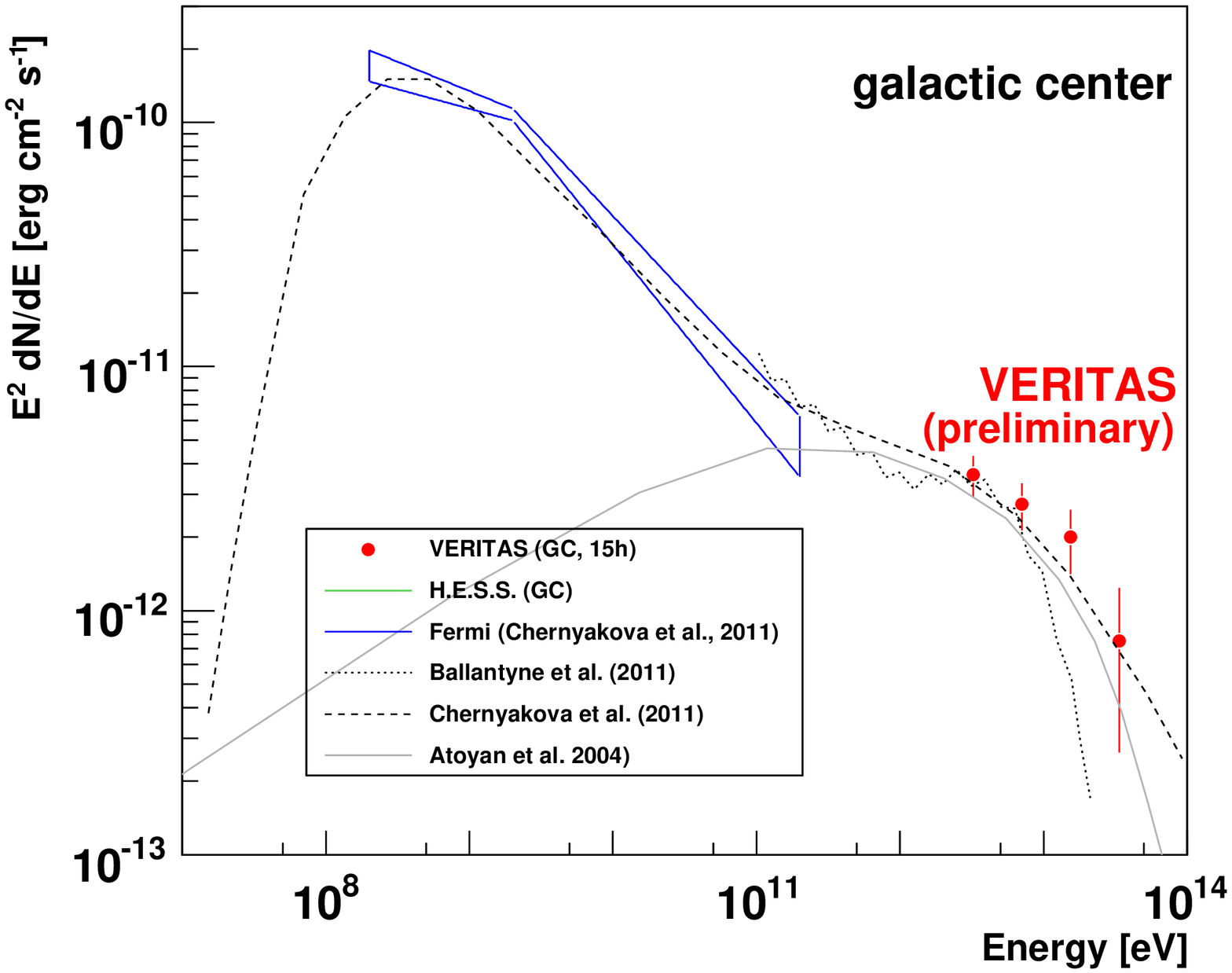}

\caption{\label{fig:SED} {\bf Left:} VERITAS energy spectrum measured
from the direction of the GC (statistical errors only). Also shown are
bow ties representing the spectra measured by Whipple \cite{Whipple_GC},
H.E.S.S. \cite{HESS_SgrA}, and MAGIC \cite{MAGIC_GC} (see
Sec.~\ref{sec:Introduction}). {\bf Right:} VERITAS energy spectrum
compared to hadronic \cite{Chernyakova2011, Ballantyne2011} and leptonic
\cite{Atoyan2004} emission models discussed for the GC source. The Fermi
bow tie is taken from \cite{Chernyakova2011}.}

\end{figure*}

VERITAS consists of four $12 \, \rm{m}$ diameter imaging atmospheric
Cherenkov telescopes and is located at the base camp of the Fred
Lawrence Whipple Observatory in southern Arizona at an altitude of $1280
\, \rm{m}$. VERITAS is sensitive to $\gamma$-rays in the energy range of
$100 \, \rm{GeV}$ to several tens of TeV. For observations close to
zenith a source of $10 \%$ ($1 \%$) of the strength of the Crab Nebula
can be detected at the level of $5$~s.d. in $0.5 \, \rm{hrs}$ ($26 \,
\rm{hrs}$), respectively.

The GC was observed by VERITAS in 2010 for $14.7 \, \rm{hrs}$ (good
quality data, dead-time corrected). Given the declination of the GC, the
observations were performed at LZA in the range of $z = 60.2-66.4 \deg$,
resulting in an average energy threshold of $E_{\rm{thr}} \simeq 2.5 \,
\rm{TeV}$. The shower direction and impact parameter were reconstructed
with the {\it geo}/{\it disp} method as described in Sec.~\ref{sec:LZA}. 
Other than that, the standard analysis procedure was applied with
standard cuts {\it a-priori} optimized for galactic sources. The column
density of the atmosphere changes with $1 / \cos(z)$. In a conservative
estimate, the systematic error in the energy/flux reconstruction can be
expected to scale accordingly. For the GC observations the contribution
of the systematic effect induced by the atmosphere therefore roughly
doubles as compared to low-zenith angle observations. Detailed studied
are needed for an accurate estimate; for the moment we give a
conservative value of a systematic error on the LZA flux normalization
of $\Delta \Phi / \Phi \simeq 0.4$.

%% ############################################################
\subsection{Results}

The VERITAS sky map of the GC region is shown in Fig.~\ref{fig:Skymap}.
An excess on the order of $12$~s.d. is detected. A fit of the point
spread function to the uncorrelated excess map results in a position of
the excess of $\rm{long} = (-0.06 \pm 0.02) \deg$ and $\rm{lat} = (-0.06
\pm 0.01) \deg$ which is well compatible with the GC position
($\rm{long} = -0.06 \deg$ and $\rm{lat} = -0.05 \deg$) and the position
measured by H.E.S.S. No evidence for variability was found in the data.
The energy spectrum is shown in Fig.~\ref{fig:SED} and is found to be
compatible with the spectra measured by Whipple, H.E.S.S., and MAGIC.
Moreover, the uncertainty in the spectrum at energies $E > 2.5 \,
\rm{TeV}$ is found to be comparable to the high-energy H.E.S.S. 
measurements since the much larger LZA effective area compensates the
shorter exposure time. To gain an independent estimate of the
systematics of the energy/flux reconstruction at LZA, a $3.5 \,
\rm{hrs}$ data set taken on the Crab Nebula at LZA ($z > 55 \deg$) was
analyzed with the same method as applied for the GC. The reconstructed
Crab Nebula spectrum is found to be in reasonable agreement with the
H.E.S.S. measurements obtained from lower zenith angles.

%% ############################################################
\subsection{Comparison to models}

Astrophysical models have been put in place to explain the GeV/TeV
$\gamma$-ray emission from the vicinity of the BH. Hadronic acceleration
models \cite{Chernyakova2011, Ballantyne2011} explain the emission by
the following mechanism: (i) Hadrons are accelerated in the BH vicinity
(few tens of Schwarzschild radii). (ii) The accelerated protons diffuse
out into the interstellar medium where they (iii) produce neutral pions
which decay into GeV/TeV $\gamma$-rays: $\pi^{0} \rightarrow \gamma
\gamma$. Changes in flux can potentially be caused by changes in the BH
vicinity (e.g. accretion). The time scales of flux variations in these
models are $\sim$$10^{4} \, \rm{yr}$ at MeV/GeV energies (old flares)
and $\sim$$10 \, \rm{yr}$ at $E > 10 \, \rm{TeV}$ ('new' flares caused
by recently injected high-energy particles) \cite{Chernyakova2011}.
While not observed over the $\sim$15 year time frame of Whipple,
H.E.S.S., MAGIC, and VERITAS spectral variability can be expected in
this model for $E > 10 \, \rm{TeV}$ with the TeV spectrum softening
following an outburst \cite{Ballantyne2011}.

Atoyan et al. (2004) \cite{Atoyan2004} discuss a BH plerion model in
which a termination shock of a leptonic wind accelerates leptons to
relativistic energies which in turn produce TeV $\gamma$-rays via
inverse Compton scattering. The flux variability time scale in this
model is on the order of $T_{\rm{var}} \sim$$100 \, \rm{yr}$ and
therefore would allow to distinguish hadronic vs. leptonic models in the
case that TeV $\gamma$-ray flux variability is detected. The hadronic
and the leptonic models discussed in this section are shown together
with the VERITAS/Fermi data in Fig.~\ref{fig:SED} (right). The leptonic
model clearly fails in explaining the flux in the MeV/GeV regime.
However, this emission may well originate from a spatially different
region or mechanism than the TeV $\gamma$-ray emission since the SED
indicates a spectral break between the Fermi/LAT and VERITAS energy
regimes. The hadronic models can explain the SED by the superposition of
different flare stages. Future Fermi/VERITAS flux correlation studies,
as well as the measurements of the TeV energy cut-off and limits on the
$E > 10 \, \rm{TeV}$ variability will serve as crucial inputs for the
modeling.

%% ############################################################
\subsection{Upper limit on diffuse $\gamma$-ray emission and dark-matter
annihilation flux}

The VERITAS observations of the GC region were accompanied by OFF-source
observations of a field located in the vicinity of the GC region
(similar zenith angles and sky brightness) without a known TeV
$\gamma$-ray source. These observations can be used to study the
background acceptance throughout the field of view and will support the
estimate of a diffuse $\gamma$-ray component surrounding the position of
the GC. An upper limit of the diffuse $\gamma$-ray flux can in turn be
compared with line-of-sight integrals along the density profile $\int
\rho^{2} \rm{d}l$, in order to constrain the annihilation cross section
for a particular dark matter model, dark matter particle mass and
density profile $\rho(r)$. Due to its likely astrophysical origin the
excess at the GC itself, as well as a region along the galactic plane,
will be excluded from this analysis (work in progress).

%% ############################################################
%% ############ Summary and Conclusion
%% ############################################################
\section{Summary and conclusion}

The implementation of the {\it disp} method into the VERITAS data
analysis chain substantially improved the shower reconstruction and
sensitivity for data taken at LZA and allowed to detect the GC within $3
\, \rm{hrs}$ in $z>60 \, \deg$ observations. The energy spectrum
measured from the GC by VERITAS is found to be in agreement with earlier
measurements by H.E.S.S., MAGIC, and Whipple. Future observations to
measure the cut-off energy in the spectrum and to determine limits on
the flux variability at the highest energies will allow to constrain the
emission models discussed in the literature. An upper limit on diffuse
$\gamma$-ray emission and, in consequence, a limit on the potential
photon flux initiated by the annihilation of dark matter particles is
work in progress.

%% ############################################################
%% ############ Acknowledgements
%% ############################################################
\bigskip % extra skip inserted
\begin{acknowledgments}

The VERITAS Collaboration acknowledges support from the US Department of
Energy, the US National Science Foundation, and the Smithsonian
Institution, from NSERC in Canada, from Science Foundation Ireland (SFI
10/RFP/AST2748), and from STFC in the UK. We acknowledge the excellent
work of the technical support staff at the FLWO and at the collaborating
institutions in the construction and operation of the instrument.

\end{acknowledgments}

%% ############################################################
%% ############ Bibliography
%% ############################################################
%\begin{thebibliography}{9}   % Use for  1-9  references


\begin{thebibliography}{99} % Use for 10-99 references

\bibitem{GC_Plerion}Q.~D. Wang, F.~J. Lu, E.~V. Gotthelf, et al.,
``G359.95-0.04: an energetic pulsar candidate near Sgr\,A*'', MNRAS, 367
(2006) 937.

\bibitem{Neutralino}G. Jungman, M. Kamionkowski, and K. Griest,
``Supersymmetric dark matter'', PhR, 267 (1996) 195.

\bibitem{GammasFromNWF}L. Bergstr{\"o}m, P. Ullio, and J. Buckley,
``Observability of gamma rays from dark matter neutralino annihilations
in the Milky Way halo'', APh, 9 (1998) 137.

\bibitem{Egret_GC}R.~C. Hartman, D.~L. Bertsch, S.~D. Bloom, et al.,
``The Third EGRET Catalog of High-Energy Gamma-Ray Sources'', ApJS, 123
(1999) 79.

\bibitem{Fermi_FirstCatalog}A.~A. Abdo, et al., ``Fermi Large Area
Telescope First Source Catalog'', ApJS, 188 (2010) 405.

\bibitem{CANGAROO_GC}K. Tsuchiya, R. Enomoto, L.~T. Ksenofontov, et al.,
``Detection of Sub-TeV Gamma Rays from the Galactic Center Direction by
CANGAROO-II'', ApJ, 606 (2004) L115.

\bibitem{Whipple_GC}K. Kosack, H.~M. Badran, I.~H. Bond, et al., ``TeV
Gamma-Ray Observations of the Galactic Center'', ApJ, 608 (2004) 97.

\bibitem{HESS_SgrA}F.~A. Aharonian, et al., ``Very high energy gamma
rays from the direction of Sagittarius\,A*'', A\&A, 425 (2004) L13.

\bibitem{HESS_SgrA_Diffuse}F.~A. Aharonian, et al., ``Discovery of
very-high-energy $\gamma$-rays from the Galactic Centre ridge'', Nature,
439 (2006) 695.

\bibitem{MAGIC_GC}J. Albert, E. Aliu, H. Anderhub, et al., ``Observation
of Gamma Rays from the Galactic Center with the MAGIC Telescope'', ApJ,
638 (2006) L101.

\bibitem{Hofmann1999}W. Hofmann, et al., ``Comparison of techniques to
reconstruct VHE gamma-ray showers from multiple stereoscopic Cherenkov
images'', APh, 12 (1999) 135.

\bibitem{BuckelyDisp}J.~H. Buckley, C.~W. Akerlof, D.~A. Carter-Lewis,
et al., ``Constraints on cosmic-ray origin from TeV gamma-ray
observations of supernova remnants'', A\&A, 329 (1998) 639

\bibitem{Chernyakova2011}M. Chernyakova, D. Malyshev, F.~A. Aharonian,
R.~M. Crocker, and D.~I. Jones, ``The High-energy, Arcminute-scale
Galactic Center Gamma-ray Source'', ApJ, 726 (2011) 60.

\bibitem{Ballantyne2011}D.~R. Ballantyne, M. Schumann, and B. Ford,
``Modelling the time-dependence of the TeV $\gamma$-ray source at the
Galactic Centre'', MNRAS, 410 (2011) 1521.

\bibitem{Atoyan2004}A. Atoyan, and C.~D. Dermer, ``TeV Emission from
the Galactic Center Black Hole Plerion'', ApJ, 617 (2004) L123.

%\bibitem{HESS_Crab}F.~A. Aharonian, et al., ``Observations of the Crab
%nebula with HESS'', A\&A, 457 (2006) 899-915.

\end{thebibliography}
\end{document}